# How marketing vocabulary was evolving from 2005 to 2014? An illustrative application of statistical methods on text mining

Igor Barahona[a*], Daría Micaela Hernández[b],
Héctor Hugo Pérez-Villarreal[c]

[a*]Research Fellow - Cátedras CONACYT, México
[b]Technical University of Catalonia. Barcelona - Tech
[c]Popular Autonomous University of Puebla State. Puebla, México.

## Abstract

Here a collection of 1169 abstracts, which corresponds to articles that the Journal of Marketing Research has published from 2005 to 2014, are analysed under a novel approach. We apply several statistical methods, such as Principal Components Analysis and Correspondence Analysis to identify the way Marketing vocabulary is evolving. Similarly those articles that introduce new vocabulary are identified and the preferred words by authors are also detected.

In order to provide an easy-to-understand explanation, we provide our results graphically. A word-cloud with the most frequent words is given first. Secondly abstracts-words are represented on the factorial plane. Finally one representation of word-years allows us to detect changes on the vocabulary through the passing of time.

## 1. Introduction

The increasing complexity on worldwide business environment, which is basically characterized by the globalization of the markets; the emergence of more powerful computers, accumulated data, proliferation of real time communication channels and social networks, are radically transformed the way organizations make decisions. According with Tuner et. al (2014) the digital data generated worldwide is growing 40% each year into the next decade. It will be expanding to include not only the increasing number of persons connected to Internet, but also the number of connected devices, as smart phones, tables, TV's and vehicles. For instance, from 2013 to 2020, the accumulated data will grow by a factor of 10. This is from 4.4 trillion gigabytes to 44 trillion. Tuner et. al (2014) also estimated that in 2013, around the 22% of accumulated data is suitable for analysis through statistical methods. This makes contrast with the 35% that is expected to be reached on 2020, mostly due to the growth of data generated from embedded systems and real time applications. It draws our attention that the owners of nearly 80% of this "digital universe" will be private companies as Google, Twitter and Facebook. In this respect, countries governments will be required to introduce regulations, in order to successfully cope with issues as privacy, security and copyrights.

* Corresponding author address: Av. Insurgentes Sur 1582, Col. Crédito Constructor . Delegación Benito Juárez C.P.: 03940, México, D.F. Telephone +52 55 5322-7700. email: jihbarahonato@conacyt.mx

Given this massive growth of digital data, it is clear that effective methods are needed in order to successfully extract knowledge from data. In the case of academic environment, methods are required to effectively perform literature reviews. According with Cooper (1998), the process of literature review is understood as identifying, assessing and synthesizing the best-available empirical evidence to answer a set of research questions. The foregoing implies that an effective literature review should identify all relevant articles and exclude irrelevant ones. Traditionally, researchers used to conduct literature reviews manually, by analysing and classifying one-by-one each document. Now, traditional methods are becoming inoperative as the available amount of articles increases. New approaches, based on information technologies and statistics are need.

Given this scenario, it is clear that organizations must act promptly, if they are willing to ensure the leadership. Competitive advantages will be given to such organizations, which are able to extract useful information from big-data and make more accurate decisions. The evidence based management (EBMgt), introduced by Rousseau (2006), makes clear the importance of analysing data to improve organization’s performance. According with this author, EBMgt is the discipline which applies the scientific method principles to make better decisions. Similarly, data mining (DM) is another discipline that emerges as response of data accumulated. According with Witten & Frank (2005) the main purpose of DM is to extract useful knowledge from raw data. Text mining (TM) is a branch of the DM, which is focused on collecting, cleaning and processing text documents. Delen & Crossland (2008) introduced the application of TM on scientific texts, with the purpose of supporting researchers to conduct literature reviews, and consequently identify trends and topics on a given research subject. Although their research is focused on top journals of the management information systems, their methodology could be applied on any research field.

The main objective of this paper is the application of the general principles of text mining, in order to provide a global perspective of academic publications on the field of marketing science. A collection of abstracts was downloaded from the Journal of Marketing Research (JMR) and later, analysed through multivariate statistical methods. In order to achieve this, we organized this paper in four sections. Material and methods on next section are presented. The obtained results on section three are given. Finally, conclusions on section four are provided.

## 2. Materials and methods

### *2.1 Data collection*

Over the years, the perspective of marketing discipline has changed in terms of focus, emphasis and priorities. With this respect, the Journal of Marketing Research (JMR) has been pioneer in introducing these changes, and therefore gathering the attention of academics, businessmen and practitioners. A collection of 1,169 abstracts, which cover the period from 2005 to 2014 were obtained from the JMR website. As additional measures of standardization, all abstracts included title, name of the first author, country, university and year of publication. On figure 1A a classification of the documents based on the country, is provided. Similarly, on figure 1B the same group of abstracts is classified according with the year of publication.

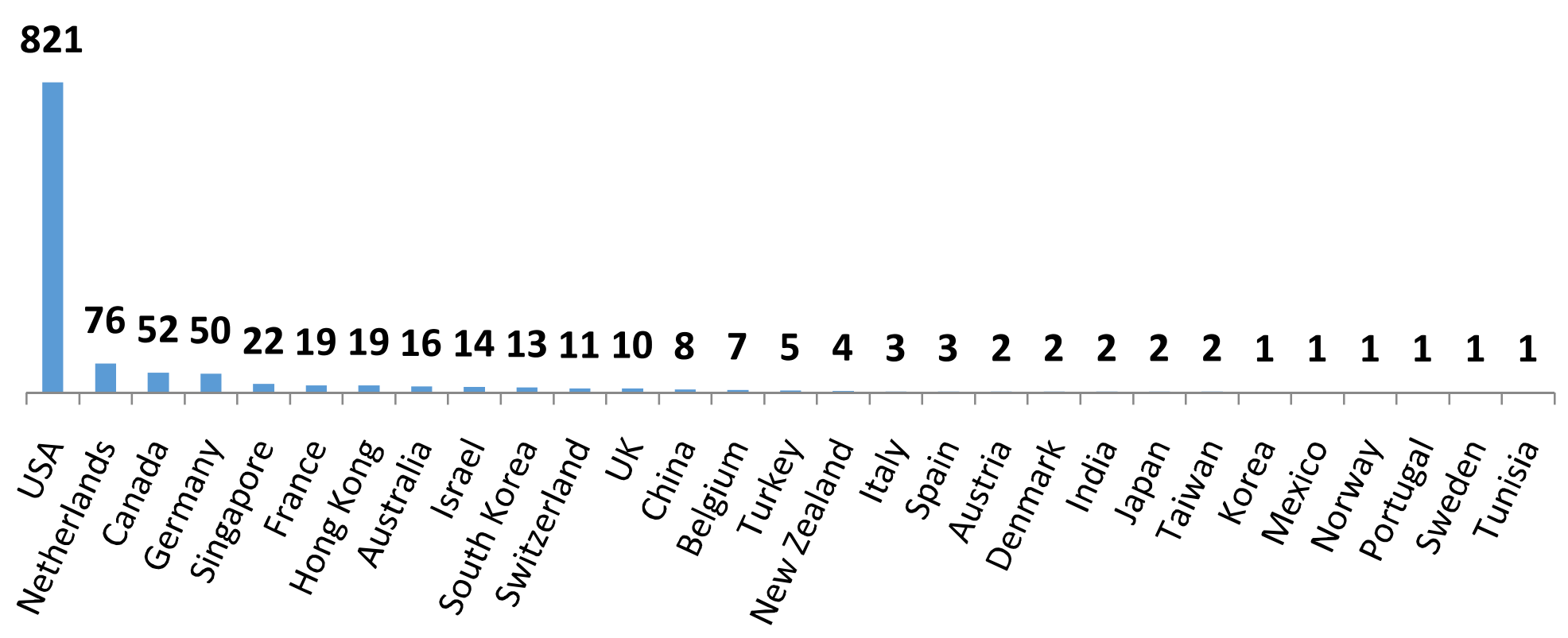

**Figure 1A**. Published articles by country from 2005 to 2014. Journal of Marketing Research.

As it is shown on figure 1A, the 70% of the publications on the JMR between 2005 and 2014 were submitted by US authors. On second place, Netherlands gathers the 6.5% of the publications. Canada is on third place with the 4.4%. Researchers from these three countries gather 81% of the all papers published by the journal. The remaining 19% is distributed among 26 different countries.

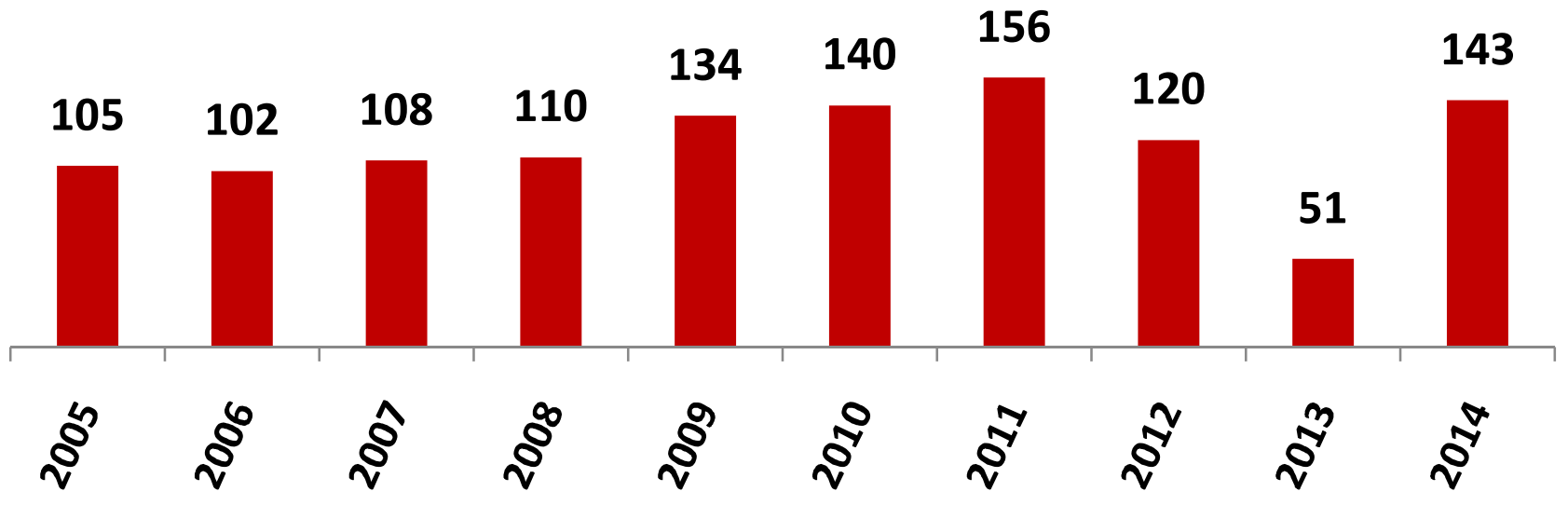

**Figure 1B**. Published articles by year. From 2005 to 2014. Journal of Marketing Research.

With respect to the year of publication, on 2011 was published the highest number of articles, equals to 156. In contrast, 2013 has the lowest number of publications, equals to 51. In short, between 2005 and 2014, the journal published on average 116 per year, with a standard deviation of 28.1.

### *2.2 Features of the dataset*

The corpus under analysis includes 1,169 documents and 120,340 terms. The foregoing represents that each abstract contains on average 103 terms. Regarding with the total number of words, the corpus has 185,437 words, which is equal to 158 words on average for each abstract. This last indicator is relevant because it lets us know the usual length of abstracts, which is accustomed by researchers who publish on the JMR.

Prior the calculation of the basic descriptive statistics, it was required to prepare the dataset. In this respect, prepositions, conjunctions, personal pronouns, articles and demonstratives, were removed. With the purpose of simplifying the analysis, words were reduced to their root form. This process is also named stemming and it is commonly used on text mining. A dictionary was specially designed for stemming purposes of this research, by applying the following logic: The words “accountability”, “accountable”, and “accounted” would be treated as “account”. Our dictionary simplified the analysis while the root meaning is preserved. This document is available upon request to the corresponding author.

**Table 1**. Descriptive statistics of the dataset under analysis.

| Descriptive Statistics | Abstract mean | Total |
|---|---|---|
| Number of terms | 102.9 | 120,340.0 |
| Number of unique terms | 71.0 | 8,874.0 |
| Percent of unique terms | 70.4 | 7.4 |
| Number of words | 158.6 | 185,437.0 |
| Average word length | 5.9 | 5.9 |

On table 1 is shown the percentage of unique terms. While the whole dataset has 7.4% percent of unique terms, the mean per abstract is equal to 70.4%. The low percentage of all documents is a measure of the vocabulary consistency. As it is stated by Bécue-Bertaut (2014), the percentage of unique terms is on inverse relation with the uniformity of the vocabulary on a given document. Thus, percentages closer to 1 indicate high diversity on the employed vocabulary. In this particular case, it is clear that the whole dataset has high vocabulary uniformity. This makes sense, considering that all abstracts were published on the same journal, and therefore they show similar features.

*2.3 Correspondence analysis*

According with Benzécri (1979), Murtagh (2005) and Bécue-Bertaut (2014) the correspondence analysis (CA) is widely used in the field of texts mining. Prior the CA, it is required to prepare our dataset and it have to be coded into a lexical table. While each abstract is assigned on a row, each unique term is allocated on a column. Every *row-column* represents the frequency for every *abstract-word* and consequently, the CA allows us to create visual representations of this lexical table, based on its frequencies. The weights for *row-abstracts* and *column-words* are automatically handled during the CA analysis. Similarly, rows and columns spaces, which are given by the chi-square distances, are used for computing the principal axes. As result, abstracts are mapped closer when they use a similar vocabulary, and words are mapped closer when they are more frequently on a given abstract. This demonstrates the suitability of CA for retrieving synonym-relationships without introducing external information on the dataset. Abstracts, whose content is closer in meaning, but expressed on different words, will be plotted closer.

On figure 2 the structure of the lexical table is shown. It has 8,874 columns, which correspond to the same number of unique terms. On the other hand, it contains 1,169 rows, that is to say one abstract per row. In addition, there are three categorical variables on the dataset that are related to year of publication, author name and university. These categorical will be incorporated to the analysis in the form of multiple correspondence analysis for mixed data.

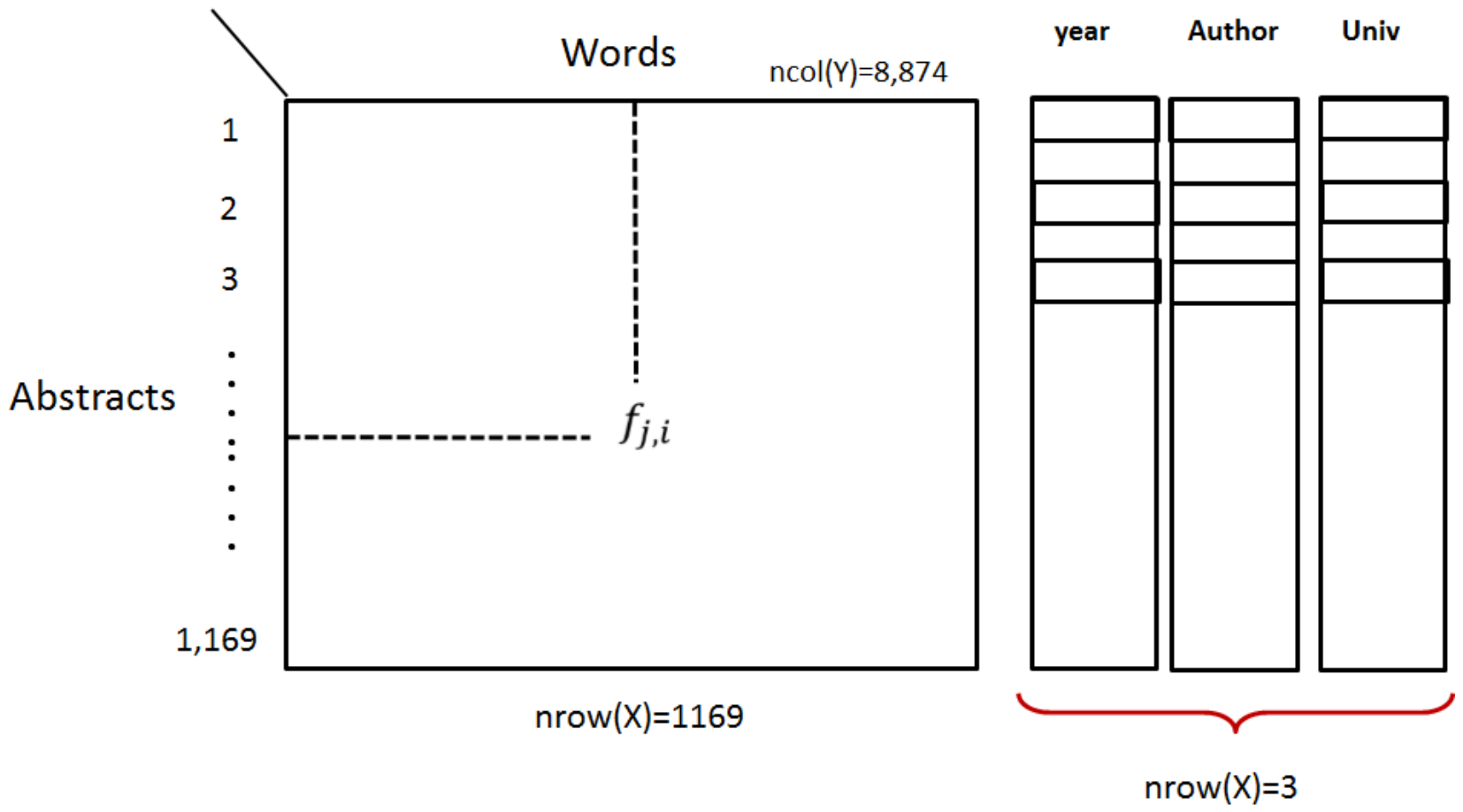

**Figure 2**. Structure of the lexical table.

By working on our lexical table through the CA, it is possible to plot abstracts and words in such a way, that hidden relations are uncovered. For instance, the identification of similarities and differences among abstracts, in terms of the employed vocabulary. Below a brief list with the outputs, which are achieved through correspondence analysis.

- Identify similarities between abstracts, given their verbal contents.
- Detect similar words, based their distribution.
- Make associations of similar words, given the context on which they were used.
- Visual representations of abstracts and words.

In terms of its interpretation, the CA attempts to seek on each axis, the “*metakeys*” and “*metadocs*”, which better describe similarities among abstracts based on the words they use. In this respect, it is important to clarify that a “*metakey*” is related with a given word, which is used one or more abstracts. On the other hand, a “*metadoc*” is related with an abstract. In this way, two or more “*metadocs*” might be related in function the same “*metakyes*”. According with Bécue-Bertaut (2014), a successful interpretation of the CA is done by mainly analysing the related the ‘‘*metakeys*’’ and ‘‘*metadocs*’’. That is to say, for each axis, the researcher might identify the set of words (‘‘*metakey+*’’/‘‘*metakey-*’’) which most contribute to the inertia and lie on the positive/negative part of the axis. Simultaneously, the set of documents that most contribute to the inertia (‘‘*metadoc+*’’/‘‘*metadoc-*’’), and lie on its positive/negative part, might be also identified.

Bansard, Kerbaol, & Coatrieux (2006) state that two “*metakeys*” or two “*metadocs*” might characterize each axis depending on its word-abstract configuration. The foregoing based on the highly contributory word-abstracts which lie at only one axis. In contrast, words frequently used in the same abstract, are all together building a topic, and they be considered to belong to the same “*metakey*”. It is important to explain that one word can belong to one or more “*metakeys*”. This scenario tells us that the same word can be used in several contexts and each context might have different meaning. For instance, the word "*environment*", may be related with the level of quality on air or water, but on another context, the same word could mean conditions and settings at the workplace. Regarding with words, which show up on many "*metakeys*", they are called by Bécue-Bertaut (2014) as "high dimensional words". These words can characterize the whole dataset with respect to one topic or field.

Moreover, the CA is able to quantitatively associate a given "*metadoc*" with a "*metakey*", on which the two together characterize an axis. With this respect, we deduce that abstracts belonging to a same "*metadoc*", are using words, which in turn are associated on the same "*metakey*". We plot on the "*metakyes*" and "*metadocs*" in order to uncover abstracts related to the same topic.

*2.4 Types of results*

Through the application of the Correspondence Analysis, and its variants which make possible the inclusion of categorical variables; we are able to obtain the three types of results, as it is stated below.

- The first approach includes global results for the active rows and columns, among which are: eigenvalues, representations of *row-abstracts* and *column-words*, distance between abstracts based on the weighted sum induced by both: vocabulary (as in CA performed on the lexical table) and the standardized chronology (classical Euclidean distance).
- On third place, we are linking the lexical matrix with the year of publication (the last, as categorical variable) in order to trace the evolution of vocabulary. In addition, the rhythm of how the vocabulary evolves is also obtained. Here, we draw the attention to gaps in vocabulary, renewals, abrupt changes or lexically homogeneous periods.

*2.5 Characteristics words*

We are modelling a hypergeometric distribution (HD) in order to identify the most frequent terms in our dataset. As it is stated on the literature, a HD defines the probability of *k* successes in *n* attempts, without replacement, given a finite population of size *N*, which contains *K* successes. Taking the HD general formulation as starting point, we propose the following notation:

– $n_{..}$, the grand total, that is, the total number of occurrences in the whole dataset;

– $n_{.j.}$, the number of occurrences in part *j*;

– $n_{i..}$, the total count of word *i* in the whole corpus;

– $n_{ij}$ the count of word *i* in part *j*.

Given this, the total $n_{ij}$ of *i* word in the part *j* is contrasted with other sums. These sums are obtained with all possible samples composed of $n_j$ occurrences randomly extracted from the whole dataset without replacement. We apply this logic: if the word *i* is relatively more frequent in the part *j* than in the whole sample, that is: $n_{ij}/n_j < n_i/n_{..}$, then the p-value is calculated as it is stated on formulations (1) and (2).

1) $p_{i,j} = \sum_{x=n_{ij}}^{n_{.j}} \frac{\binom{n_{i.}}{x}\binom{n_{..}-n_{i.}}{n_{.j}-x}}{\binom{n_{..}}{n_{.j}}}$ 2) $p_{i,j} = \sum_{x=1}^{n_{ij}} \frac{\binom{n_{i.}}{x}\binom{n_{..}-n_{i.}}{n_{.j}-x}}{\binom{n_{..}}{n_{.j}}}$

*2.6 Validating the results*

By the application of formulas presented on the previous section, we apply the methodology proposed by Bécue-Bertaut (2014) and conduct a test (one-tail test) to assess the significance of the first eigenvalue and therefore, to establish a quantitative link between chronology and vocabulary. $H_o$ comprises: There is not exist a chronological dimension of the vocabulary and hence, the exchangeability of the year of publication with respect to the abstracts. Randomly, we permute the year-columns in the dataset and, the value of the test is calculated on every permutation. As many values of the test are acquired, we build an empirical distribution of the first eigenvalue under $H_o$. It is important to conduct a large number of replications in order to compute an accurate p-value.

*2.7 Chronological words*

We might have the case on which a set of words better characterize successive periods rather a unique period. Under this scenario, each word should be tested, for a given period. Secondly, for two consecutive periods and thirdly for a group of three periods should be also tested. This process allow us to label words on which, it is failed to accept the null hypothesis as “chronological characteristics words”. That is to say, the group of periods with the lowest p-value (taking as cut-off a value of 0.05) are considered as chronological words and therefore, they allow us to comprehend how the vocabulary has been evolving.

*2.7 The statistical software*

Our statistical calculations were carried out under the software R version version 3.1.2 (2014-10-31) - "Pumpkin Helmet". The function BiblioMineR, developed by Hernández (2012), was utilized in all our analysis. This function was later incorporated to the package TextoMineR, which is currently under pilot testing.

## 3. Results

*3.1 Glossary of most frequent terms*

First, we analyse our results based on word frequencies. A total of 1,169 abstracts and 8,874 unique words, were included in our calculations. On table 2, the 25 most frequent words are shown. On the other hand, we used the 377 most frequent terms to draw a wordcloud which is presented on figure 3.

The word “*authors*” was the most frequent word with 1,850 occurrences, which means that it appears on every 1.60 abstracts in average. The second place was “*effects*”, followed by “*marketing*”. The words “*Product*” and “*Consumers*” were on fourth and fifth place respectively. On the other hand, as it's shown on figure 3, all marketing related words are more evenly distributed. Since only articles published on the Journal of Marketing Research (JMR) are here analysed, it makes sense the word "*authors*" (1850) is at the top of the list. Similarly, it is expected to have the word "*effects*" (1237) on second place, followed by "*marketing*" (1076). These three terms together make up the main topic on the journal: "*Authors discuss effects of marketing*". The foregoing provides quantitative evidence that the JMR is centred on its aims and scopes (detailed information about the journal can be reached at http://journals.ama.org/loi/jmkr).

A second group is integrated by the words "*product*" (1011), "*consumers*" (972), "*brand*" (884), and "*customer*" (844). These four terms are shaped the main contents of the journal. We identified that nearly the 28% of the abstracts are including these four words together. This overall perspective allows us to take a first approach on what seems to be an interest for JMR authors. Their efforts are directed to discuss effects on products, consumers, brand and consumers through the "*data*" (577), "*model*" (575), "*study*" (542), "*market*" (541) and "*results*" (543). In essence, the main topics of the JMR can synthesized as it was described.

With respect to figure 3, it is clear that it supports what it was stated before by highlighting the relation "*authors*" → "*effects*" → "*marketing*". The main topics of the journal are revealed on figure 3, which are also related with the words presented on the table 2.

*3.2 Correspondence analysis of the "Abstracts-words"*

The initial lexical table is of the order 1169 x 8874. The foregoing represents that it includes 1169 *row-abstracts* and 8874 *columns-words*. Performing a correspondence analysis on a large dataset just like this, might be confusing due to the majority of the elements don't contribute significantly to the analysis. A practical solution to this problem is proposed by Meyer, Hornik & Feinerer (2008), who provide a method to remove sparse terms. This treatment dramatically reduces the size, without losing significant relations inherent to the "*abstracts-words*" matrix. We filter out the words and retain only the most frequent ones. The function *remove Sparse Terms (DTM, S)*, available on the tm package is here applied. The selected sparse parameter was 0.9631 and it retains the 377 most frequent words, which are also the same words displayed on figure 3. After the sparse process was conducted, the lexical table was reduced to the form 8874 to 377.

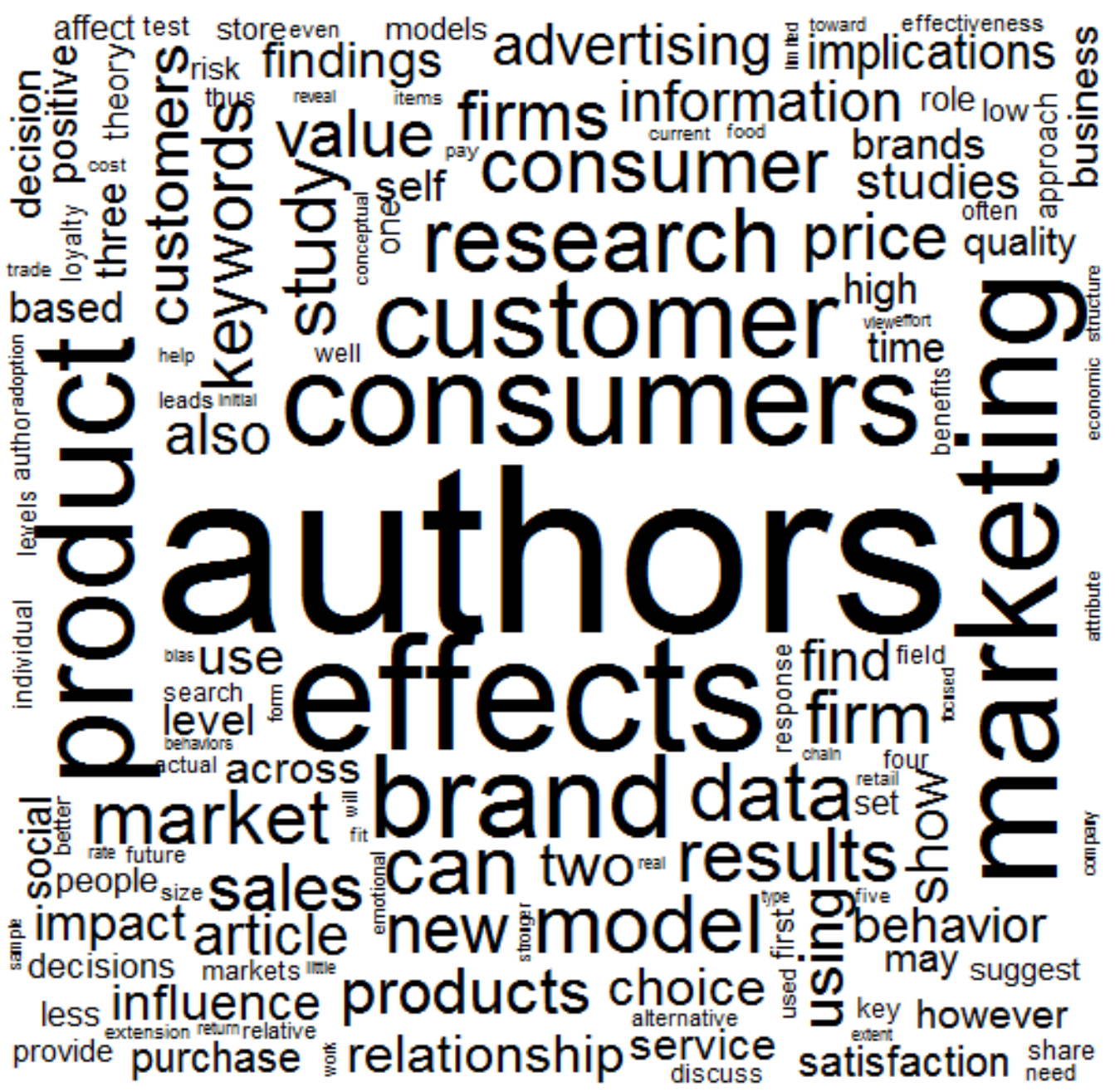

**Figure 3**. Wordcloud with the 380 most frequent terms

**Table 2**. List of the 25 most frequent terms

| Word | Frequency | Num of Abstracts |
|---|---|---|
| authors | 1850 | 918 |
| effects | 1237 | 696 |
| marketing | 1076 | 442 |
| product | 1011 | 361 |
| consumers | 972 | 416 |
| brand | 884 | 212 |
| customer | 844 | 232 |
| research | 657 | 462 |
| can | 588 | 415 |
| data | 577 | 394 |
| model | 575 | 305 |
| consumer | 554 | 331 |
| study | 542 | 334 |
| market | 541 | 240 |
| results | 534 | 431 |
| firm | 530 | 210 |
| new | 493 | 242 |
| firms | 489 | 255 |
| keywords | 482 | 475 |
| sales | 476 | 160 |
| price | 475 | 149 |
| value | 463 | 182 |
| performance | 455 | 193 |
| products | 439 | 211 |
| using | 434 | 362 |

At this point, we are ready to perform the correspondence analysis and obtain the *metakeys* and *metadocs* related with the principal axes. On table 3, the captured variance on the first 9 axes is shown. The first two axes concentrate 18.23% and 14.34% of the variance respectively, and thus they are gathering the highest number of contributory abstracts and words.

**Table 3**. Extracted variance of the first 10 axes.

| Descriptive | Dim.1 | Dim.2 | Dim.3 | Dim.4 | Dim.5 | Dim.6 | Dim.7 | Dim.8 | Dim.9 |
|---|---|---|---|---|---|---|---|---|---|
| Variance | 0.032 | 0.026 | 0.023 | 0.020 | 0.020 | 0.017 | 0.015 | 0.013 | 0.013 |
| % | 18.23 | 14.34 | 12.70 | 11.46 | 11.00 | 9.37 | 8.61 | 7.25 | 7.05 |
| Cumulative | 18.23 | 32.57 | 45.27 | 56.72 | 67.72 | 77.09 | 85.70 | 92.95 | 100.00 |

We proceed to represent the most contributory *abstracts-words* on the first two axes. Words with frequency lower than 53 were removed from analysis. Similarly, documents which use less than 15 times one word, were also eliminated. On figure 4, the visual representation of *abstracts-words* over the factorial plane.

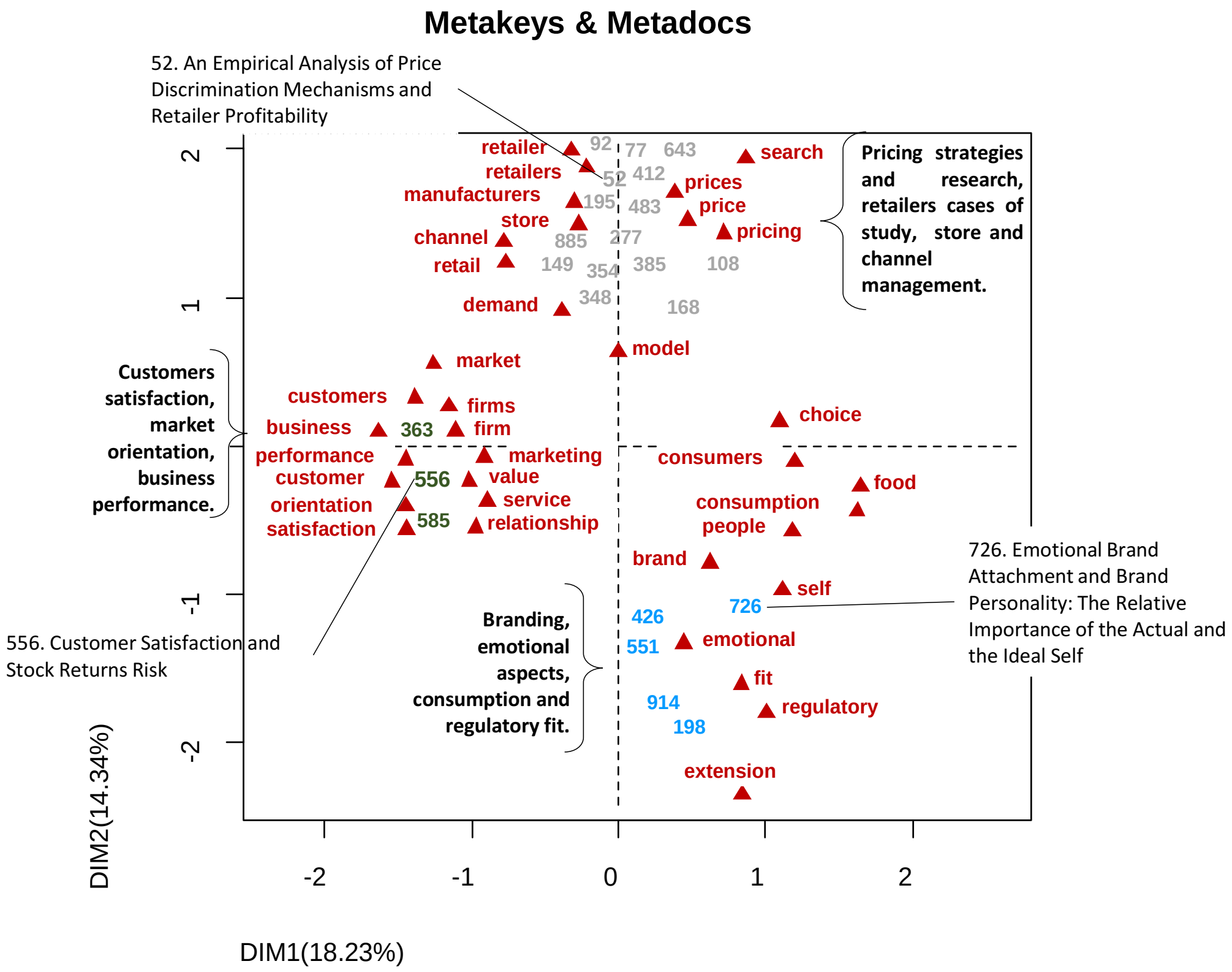

**Figure 4**. Most contributory abstracts / words on the factorial plane.

These were the criteria to select the abstract-words that are represented in figure 4: All *metadocs* with a contribution 8 times greater than average contribution were kept. Similarly, *metakeys* 5 times greater than the mean contribution were retained. Three different groups were revealed through the correspondence analysis. On the first place with grey numbers, a group of abstracts which are discussing topics as pricing strategies, channels, store management, and cases of study about retailers. A total of 15 abstracts are representative of these topics. On second place, there are three abstracts which strongly relate with topics as customer satisfaction, market orientation and business performance. They are indicated with green numbers. Finally, a set of 5 abstracts, indicated in blue, which characterize topics as branding, emotional aspects, consumption and regulatory issues.

This correspondence analysis allow us to state that there are 23 abstracts, which best characterize the aim and scope of the Journal. They are also separated on the three groups of topics, as it was previously explained.

### *3.3 Chronological evolution*

In order to find chronological patterns on our dataset, we are using the proposed by Kostov et. al (2013), which incorporates the package FactoMineR developed by Lê, Josse, & Husson (2008).

As it will be explained below, the following questions are answered: Which groups, among year of publication and vocabulary, are either similar or different? Given the groups of years, which are the similarities among the used vocabulary? And how the vocabulary has evolved over the years?

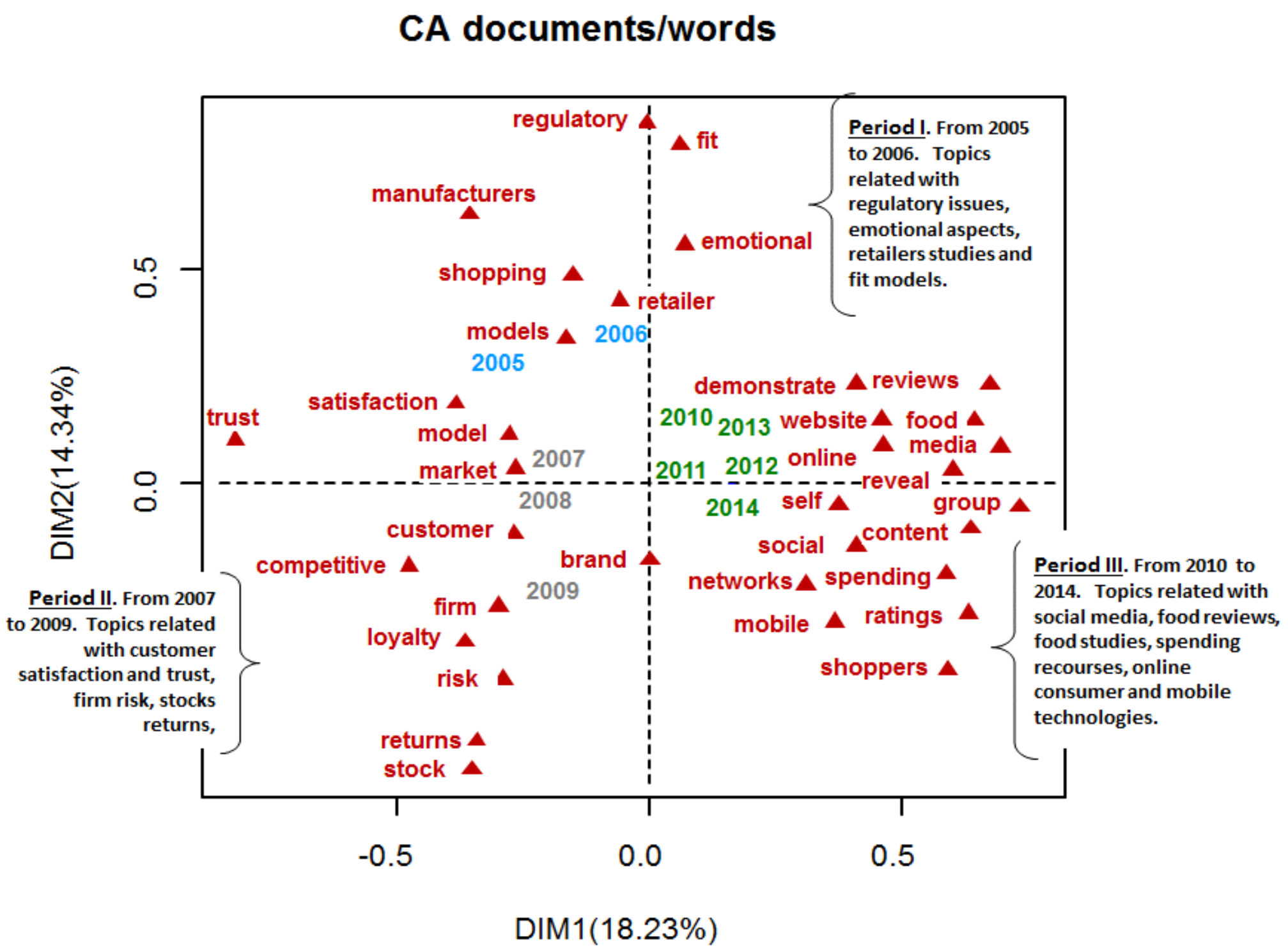

**Figure 5**. Visual representation of most frequent words and the year of publication

On figure 5 the CA for the most contributory words and theirs the year of publication, are shown. We identify three periods the first (2005-2006 in blue; 2007-2009 in grey and 2010-2014 in green). For the horizontal axis, while words related with customer satisfaction and market model on the negative part of the axis are displayed; words referring to social networks and mobile technologies are projected on the positive part. With respect to the vertical axis, the positive part is characterized by the words regulatory, fit and retailer. On its negative part, brand, networks, demonstrate and stock, are found.

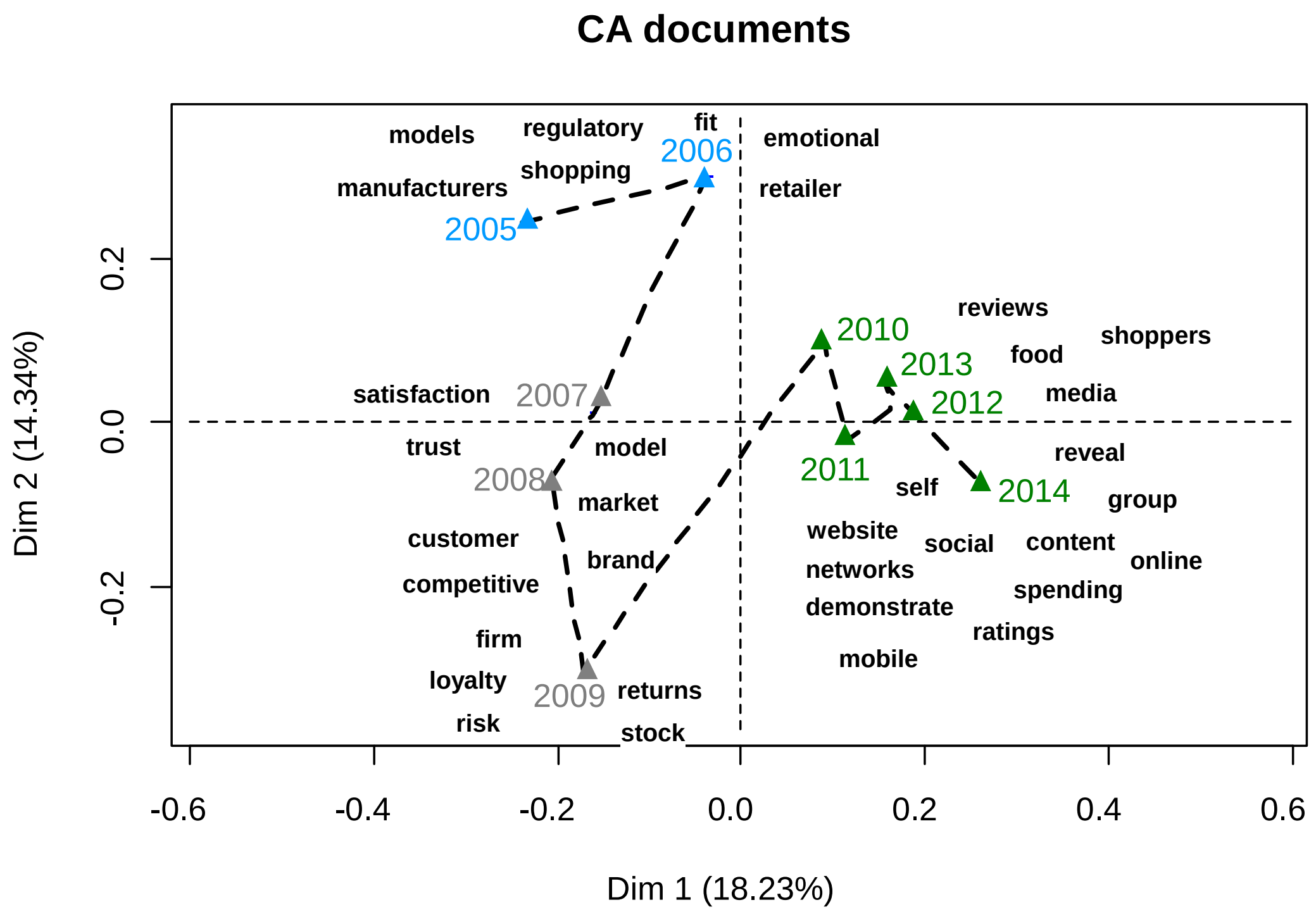

**Figure 6**. Visual representation of the evolution of the vocabulary in the first CA plane.

On figure 6 the way the vocabulary has evolved is shown. On the first period, from 2005-2006 authors were mainly writing about regulatory issues, emotional shopping and fitting models. In contrast, during the second period authors had drawn their attention to the satisfaction, loyalty and trust of the customer. Likewise, topics as market models, branding, firm returns and stocks are characteristic of this period. Finally on the third period, which comprises the years 2010 to 2014, emerged topics as social networks and contents, mobile technologies, online shoppers, reviews and demonstrations. It draws our attention the radical change that is observed on the emerged words of the period three, in contrast with the previous ones. It is evident that authors focused their attention on the contemporary tendencies, as proliferation of the marketing online and social networks.

## 4. Conclusions

On this research, an exhaustive collection of 1169 abstracts, which cover a 9 years period, were investigating by implementing a novel statistical methodology. All abstracts correspond to articles that the Journal of Marketing Research has published. By way of introduction, descriptive statistics such as, the words average per abstract, percentage of unique terms and average word length, were discussed. Thereafter, the most frequent words were identified and a wordcloud allowed us to disclose the preferred vocabulary by authors. By conducting a correspondence analysis the most influential abstracts were identified. Finally, a multifactor analysis of contingency tables was calculated to disclose how the use of vocabulary has been evolving. With this respect, three main periods were categorized.

This article provides strong evidence of the hot topics in the marketing science. In the same way, these results might be used as guideline by authors and researchers who are interested in submitting their research to the JMR. That is to say, by comprehending the specialized vocabulary that is used on this discipline, and later incorporating it on their documents, authors might assure that they are at the forefront of vocabulary usage.